\documentclass[twocolumn]{aastex631}

\usepackage{amsmath}
\usepackage{mathrsfs}
\usepackage{wasysym}
\usepackage{hyperref}
\usepackage{bm}

\newcommand{\refeq}[1]{Eq.~(\ref{eq:#1})}          
          
\newcommand{\reffig}[1]{Figure~\ref{fig:#1}}

\newcommand{\reftab}[1]{Table~\ref{tab:#1}}

\begin{document}

\title{Under Einstein's Microscope: Measuring Properties of Individual Rotating Massive Stars from Extragalactic Microcaustic Crossings}

\author{Xu Han}
\affiliation{Department of Physics, Xi'an Jiaotong University, Xi'an, 710049, China}
\affiliation{Department of Physics, University of California, 366 Physics North MC 7300, Berkeley, CA 94720, USA}

\author[0000-0003-2091-8946]{Liang Dai}
\affiliation{Department of Physics, University of California, 366 Physics North MC 7300, Berkeley, CA 94720, USA}
 
\begin{abstract}

Highly magnified stars residing in caustic crossing lensed galaxies at $z\simeq 0.7$--$1.5$ in galaxy cluster lensing fields inevitably exhibit recurrent brightening events as they traverse a microcaustic network cast down by foreground intracluster stars. The detectable ones belong to nature's most massive and luminous class of stars, with evolved blue supergiants being the brightest ones at optical wavelengths. Considering single stars in this work, we study to what extent intrinsic stellar parameters are measurable from multifilter light curves, which can be obtained with optical/near-IR space telescopes during one or multiple caustic crossing events. We adopt a realistic model for the axisymmetric surface brightness profiles of rotating O/B stars and develop a numerical lensing code that treats finite source size effects. With a single microcaustic crossing, the ratio of the surface rotation velocity to the breakup value is measurable to a precision of $\sim 0.1$--$0.2$ for feasible observation parameters with current space telescopes, with all unknown intrinsic and extrinsic parameters marginalized over and without a degeneracy with inclination. Equatorial radius and bolometric luminosity can be measured to $1/3$ and $2/3$ of the fractional uncertainty in the microcaustic strength, for which the value is not known at each crossing but an informative prior can be obtained from theory. Parameter inference precision may be further improved if multiple caustic crossing events for the same lensed star are jointly analyzed. Our results imply new opportunities to survey individual massive stars in star-formation sites at $z\simeq 0.7$--$1.5$ or beyond.

\end{abstract}


\section{Introduction} \label{sec:intro}

When a background galaxy overlaps the lensing caustic cast by a foreground galaxy cluster lens, a portion of that galaxy in the caustic vicinity appears immensely stretched. Stars residing in that region can be magnified by a few hundred to a thousand fold, and the most luminous ones are individually detectable~\citep{1991ApJ...379...94M}. A significant advance in our understanding of highly magnified stars is the realization that intracluster stars in the foreground cluster inevitably disrupt the smooth cluster caustic into a network of microcaustics on a fine scale $\sim 10^3\,$au~\citep{Venumadhav2017CausticMicrolensing, Diego2018DMUnderMicroscope, Oguri2018CausticCrossing}. Highly magnified stars therefore appear variable as they traverse the nonuniform magnification pattern on the source plane as a result of the microcaustic network.

Highly magnified stars can be detected based on the variability induced by microlensing. Without microlensing, they are expected to be common at $z=1$--$2$ in deep imaging that reaches $28$--$29$ mag in the optical and near-IR~\citep{Diego2019ExtremeMagnification}. Microlensing can further boost the flux by $1$--$3$ mag temporarily, which greatly enhances detectability. Several blue supergiant stars have already been discovered by the Hubble Space Telescope (HST) in caustic-crossing giant arcs at $z=0.7$--$1.5$~\citep{Kelly2018NatAsM1149, Chen2019LensedStarM0416, Kaurov2019LensedStarM0416}. Searching for variability using the imaging difference technique, the Flashlight program recently detected a dozen highly magnified stars from this redshift range when the imaging limiting magnitude with HST in the optical improves from $28$ to $30$~\citep{Kelly2022FlashlightDozenStars}. James Webb Space Telescope (JWST) NIRCam imaging with the PEARLS program uncovered that redder magnified stars are commonly detected at infrared wavelengths at a magnitude limit $28.5$--$30$~\citep{Yan2023ApJS..269...43Y}. Several intriguing magnified star candidates have been reported in more distant lensed galaxies at $z\simeq 2$--$6$~\citep{Welch2022EarendelNature, Diego2023PearlsElGordo, Diego2023Mothra, meena2023two,welch2022jwst}, including quite bright objects~\citep{vanzella2020probing, diego2022godzilla}. The nature of these sources is under active investigation.

Microlensing-induced brightening culminates every time when the source star transits a microcaustic. When the photosphere overlaps the microcaustic, a peak magnification $\mu_{\rm pk} \sim 10^3$--$10^4$ is reached over the stellar-diameter crossing time (hours to a few days). Flux evolution around such dramatic times has yet to be measured at high cadence but is observationally feasible owing to the expected immense magnification factor around the peak time.

In this work, we study the potential of caustic crossing as an exquisite gravitational microscope to measure stellar properties. During the caustic crossing, different parts of the stellar surface have significantly different magnification factors, and thus the time-dependent fluxes and colors encode information about the size of the photosphere, as well as its shape and surface temperature profile. 

This can be compared to the study of finite source size effects in Galactic binary microlensing events with caustic crossings~\citep{MaoPaczynski1991BinaryMicrolensing}, with some observed examples~\citep{Lennon1996CausticCrossing, OGLE:2003zjz, PLANETRoboNet:2005kev}. In extragalatic caustic crossings it is the most massive stars  with the highest luminosities that are observationally selected. Therefore, caustic crossings with highly magnified stars provide a unique opportunity to study the properties of individual massive stars in distant star-formation environments that might differ fundamentally from local universe samples (in terms of, e.g., metallicity and interstellar medium density).

In this work, we will study single source stars that have axisymmetric surface properties owing to stellar rotation. We will introduce a parameterized model that accounts for the axisymmetric surface spectral energy distribution (SED) profile due to gravity darkening. By resorting to analytic arguments and by developing a numerical program to fit mock multifilter light curves for caustic-crossing events, we will show that, with all unknown extrinsic geometric and kinematic parameters marginalized over, the ratio of the surface rotation velocity to the breakup velocity (see~\refeq{omega}), denoted as $\omega$, can be measured without a degeneracy with the stellar inclination as in spectroscopic rotation measurements. For massive stars, the evolution of stellar rotation connects to wind mass loss~\citep{MeaderMeynet2000CriticalRotationVSEddington}, surface element abundance enhancement (e.g. N and He;~\cite{MaederMeynet2000RotatingStarsARAAReview}), and the explosive fate of the star~\citep{WoosleyHeger2006GRBprogenitors, Yoon2006LongGRBs}. Knowing $\omega$ for a sample of massive stars at $z=0.7$--$1.5$, combined with the knowledge of the property of each star's host environment, can potentially improve our understanding of massive star evolution in the younger Universe. 

Additionally, the equatorial radius $R_e$ and the bolometric luminosity $L$ can be constrained to precision levels limited primarily by the unknown microcaustic strength (see \refeq{dstar}) the logarithmic uncertainties (in units of dex) for $R_e$ and $L$ are $1/3$ and $2/3$ of that for the microcaustic strength, respectively. Fortunately, an informative prior for the latter can be derived from microlensing simulations knowing their number density and mass distribution~\citep{Venumadhav2017CausticMicrolensing}.

This paper will be organized as follows. In Sec.~\ref{sec2}, we discuss an analytical model that quantifies the axisymmetric surface brightness profile of a rotating massive star and introduce intrinsic and extrinsic parameters of the model. In Sec.~\ref{sec3}, we summarize how individual microcaustic crossing events can be modeled by a local fold lens model and define the relevant parameters. In Sec.~\ref{sec4}, we discuss parameter degeneracy expected in fitting mock multifilter light curves during the caustic-crossing process to our theoretical model. In Sec.~\ref{sec5}, we present examples of mock parameter inference under realistic observational conditions, which we carry out using a numerical code we have developed that performs inverse ray shooting and rapidly calculates the light-curve fitting $\chi^2$ as a function of model parameters. Sec.~\ref{sec6} is for further discussions, and we will close with concluding remarks in Sec.~\ref{sec7}.

\section{Source Star Model}
\label{sec2}

In this section, we discuss how we model the surface brightness profile of axisymmetric rotating stars.

\subsection{Surface Property of Rotating Stars}
\label{sec:Gravi-darkening}

We consider general stars with surface rotation. The surface rotation angular velocity may not equal that of the stellar interior. The surface of a rotating star has an axisymmetric shape and has a latitudinal effective temperature profile $T_{\rm eff}(\theta)$ (where $\theta$ is the polar angle and is equal to $\pi/2$ minus the latitude). The von Zeipel theorem~\citep{vonZeipel1924} states that the bolometric radiation flux $F$ emerging from the surface of the rotating star is proportional to the local effective surface gravity $g_{\rm eff}$, which is centrifugally reduced on the equator. As a result, the polar caps are hotter than the equatorial region and have a higher surface brightness.

We adopt the analytic model of \cite{Lara2011GravityDarkening} for the axisymmetric distribution of surface radiation flux $F(\tilde{r},\,\theta,\,\omega)$ and surface gravity $g_{\mathrm{eff}}(\tilde{r},\,\theta,\,\omega)$:
\begin{align}
\label{eq:F}
    F & = \sigma\,T_{\mathrm{eff}}^4 = {g_{\mathrm{eff}}}\,\left(\frac{\tan{\vartheta}}{\tan\theta}\right)^2 \frac{L}{4\pi\,G M}, \\
\label{eq:geff}
    g_{\mathrm{eff}} & = \frac{GM}{R_e^2} \left(\frac{1}{\tilde{r}^4}+\omega^4\,\tilde{r}^2\,\sin^2\theta-\frac{2\,\omega^2\sin^2\theta}{\tilde{r}}\right)^{\frac12}.
\end{align}
More sophisticated than the simple power-law model $T_{\rm eff} \propto g^{\beta}_{\rm eff}$ with $\beta\approx 1/4$ for massive stars with a radiative envelope, this model agrees well with 2D numerical rotating star models. In this model, the reduced radius $\tilde{r}$ is defined as the dimensionless ratio of the radial position of the stellar surface to the equatorial radius $R_e$, i.e. $\tilde{r} \equiv r/R_e$. This quantifies the axisymmetric deformation of the surface due to the centrifugal force. The quantities $L$ and $M$ are the bolometric luminosity and the stellar mass, respectively, and $\sigma$ is the Stefan-Boltzmann constant. Following \cite{Lara2011GravityDarkening}, we have introduced an auxiliary variable $\vartheta$, which is related to the polar angle $\theta$ via \refeq{shape}.

The dimensionless rotation parameter $\omega$ is a measure of how dynamically important the surface rotation is. It is defined as
\begin{equation}
\label{eq:omega}
    \omega \equiv \Omega \left(\frac{R_e^3}{GM}\right)^{\frac12}= \frac{\Omega}{\Omega_k}.
\end{equation}
Here $\Omega$ is the angular velocity of the surface, and $\Omega_k$ is the Keplerian angular velocity on the equatorial surface, which is the critical rotating rate for material on the equatorial surface to be bound in a circular orbit. The model requires $0\leqslant \omega < 1$. If the value is close to breakup $\omega \gtrsim 0.7$--$0.8$, a decretion disk is likely to form on the equator to form a Be star~\citep{Ekstrom2011IAUS..272...62E}.

The following implicit equation gives an analytic relation between the polar angle $\theta$ and $\vartheta$:
\begin{equation}
\label{eq:shape}
    \cos\vartheta+\ln\left(\tan\frac{\vartheta}{2}\right) = \frac{1}{3}\,\omega^2\,\tilde{r}^3\,\cos^3\theta + \cos\theta + \ln\left(\tan\frac{\theta}{2}\right).
\end{equation}
The variable $\vartheta$ reduces to $\theta$ in the absence of rotation.

The shape of the star $r(\theta)$ can be found from a constant effective potential on the surface, whose value can be determined by considering the equator:
\begin{equation}
    \frac{GM}{r}+\frac{1}{2}\,\Omega^2\,r^2\,\sin^2\theta = \frac{GM}{R_e} + \frac12\,\Omega^2\,R^2_e.
\end{equation}
It is assumed here that the gravitational potential at the stellar surface is well approximated by $GM/r$ (i.e. the Roche potential), as the mass distribution of a massive star is rather centralized~\citep{Lara2011GravityDarkening}. Recasting the equation into the dimensionless form, we derive the shape equation
\begin{equation}
\label{eq:shapeEq}
    \frac{1}{\omega^2\,\tilde{r}}+\frac{1}{2}\,\tilde{r}^2\,\sin^2\theta = \frac{1}{\omega^2}+\frac{1}{2}.
\end{equation}

The sequential procedure to compute the rotating star model is the following. We find $\tilde r(\theta, \omega)$ by solving \refeq{shapeEq} and then find $\vartheta(\theta, \omega)=\vartheta(\tilde{r}(\theta, \omega), \theta, \omega)$, and then we determine the effective surface gravity $g_{\mathrm{eff}}(\theta, \omega) = g_{\mathrm{eff}}(\tilde{r}(\theta, \omega), \theta, \omega)$ according to \refeq{geff} as well as the surface radiation flux $F(\theta, \omega) = F(g_{\rm eff}(\theta, \omega), \vartheta(\theta, \omega), \theta)$ according to \refeq{geff}. Knowing $F(\theta, \omega)$, the surface temperature $T_{\rm eff}(\theta, \omega)$ is found from $F(\theta, \omega)=\sigma\,T^4_{\rm eff}(\theta, \omega)$.

\begin{figure}[htbp]
\centering
\epsscale{1.1}
\plotone{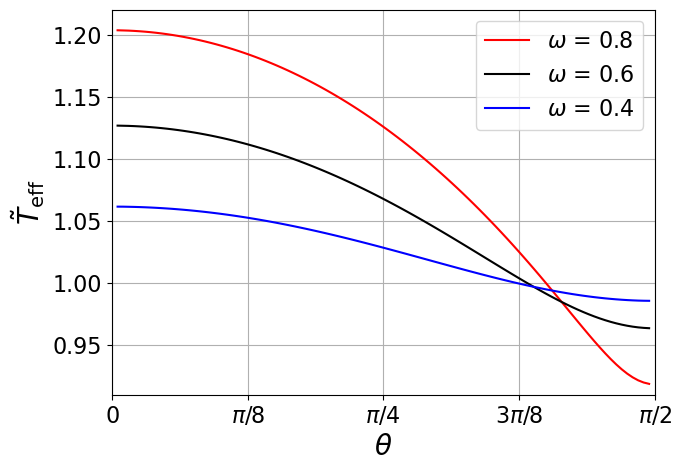} 
\plotone{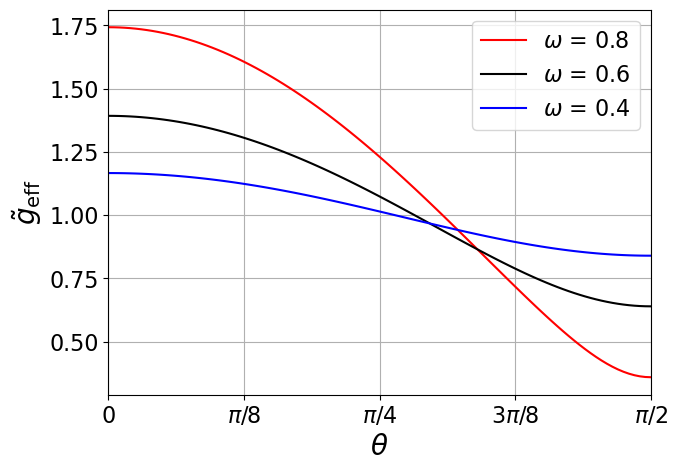} 
\caption{Reduced surface temperature $\tilde{T}_{\mathrm{eff}}(\theta, \omega)$ and reduced effective surface gravity $\tilde{g}_{\mathrm{eff}}(\theta, \omega)$ as a function of the polar angle $\theta$ for three different values of $\omega$.
\label{fig:tilde}}
\end{figure}

Defining characteristic quantities $T_{\mathrm{eff,0}} \equiv \left({L}/{4\pi\sigma R_e^2}\right)^{1/4}$ and $g_{\mathrm{eff},0} \equiv {GM}/{R_e^2}$, we can express the surface temperature and the effective surface gravity in dimensionless forms
\begin{align}
    \tilde{T}_{\mathrm{eff}}(\theta, \omega) &\equiv \frac{T_{\mathrm{eff}}(\theta, \omega)}{T_{\mathrm{eff,0}}} = \left[ \tilde{g}_{\mathrm{eff}}(\theta, \omega) \right]^\frac14 \left( \frac{\tan{\vartheta}(\theta, \omega)}{\tan\theta} \right)^{\frac12},  \\
    \Tilde{g}_{\mathrm{eff}}(\theta, \omega) &\equiv \frac{g_{\mathrm{eff}}(\theta,\omega)}{g_{\mathrm{eff},0}} \nonumber\\
    & = \left(\frac{1}{\tilde{r}^4(\theta, \omega)}+\omega^4\,\tilde{r}^2(\theta, \omega)\,\sin^2\theta-\frac{2\,\omega^2\,\sin^2\theta}{\tilde{r}(\theta, \omega)}\right)^{\frac12}.
\end{align}
In \reffig{tilde}, we show these reduced quantities as a function of the polar angle $\theta$ for a few different values of $\omega$.

\subsection{SED of Rotating Stars}

To calculate the observed SED of a lensed star, we must integrate the spectral surface brightness over the lensed image. For OB stars, the blackbody SED that depends on a single parameter $T_{\rm eff}$ would be too crude an approximation. We adopt the more realistic TLUSTY stellar spectrum templates computed for nonrotating OB stars~\citep{LanzHubeny2003tlustyOstars, LanzHubeny2007tlustyBstars}. Each model outputs the emergent SED, which is characterized by three parameters: effective temperature, surface gravity, and metallicity. 

The TLUSTY grid we use covers a range of effective temperature from $T_{\rm eff} = 15,000$ to $30,000\,\mathrm{K}$ and a range of surface gravity from $\log g=1.75$ to $4.75$. To model rotating stars, we use the approximation~\citep{collins1966theoretical, 1970A&A.....7..120M} that the SED of the local emergent radiation at any point on the surface of the rotating star is that of a nonrotating star that has an effective temperature $T_{\rm eff}$, a surface gravity $g_{\rm eff}$ and a metallicity $Z$. By assuming that radiation transfer through the stellar atmosphere is a local process, we can model the varying SED across the rotating star surface by interpolating a grid of nonrotating TLUSTY models. Limb darkening is not modeled in this work, but we will comment on that in Section \ref{sec:extmodel}.

Our approximation is imperfect, as certain spectral features depend on the radiation-hydrodynamic solution describing the line-driving wind of a massive star. The wind solution can differ between nonrotating and rotating stars. Since the relevant observable for our purpose is wide-filter photometry, wind features are not expected to be important.

For examples shown in the later part of the paper, we will choose a TLUSTY grid with a metallicity $Z=0.2\,Z_\odot$ and a microturbulence $v_{\mathrm{turb}}=2\,\mathrm{km\,s^{-1}}$. We will suppress explicit dependence on these two parameters in equations.

Note that highly magnified stars may be cool stars for which the TLUSTY models are not applicable. They may be red supergiants with $T_{\rm eff}=3500$--$4500\,$K~\citep{Diego2023PearlsElGordo}. Even the first detected star has a lower effective temperature $T_{\rm eff}=11,000$--$14,000\,$K~\citep{Kelly2018NatAsM1149}. Nevertheless, our framework is sufficiently general that any other stellar atmosphere models appropriate for different stellar types can be incorporated if needed. For a proof of concept, we will consider in this work relatively hot stars for which the TLUSTY templates are valid.

Let $\tilde S_\lambda(\lambda, T_{\rm eff},\,g)$ be the surface brightness of the TLUSTY model at wavelength $\lambda$. The surface brightness for a source star at a redshift $z_s$ observed at wavelength $\lambda$ is
\begin{align}
    S_\lambda\left(\lambda, T_{\rm eff},\,g_{\rm eff}\right) = \frac{\tilde S_\lambda\left(\lambda/(1+z_s), T_{\rm eff},\,g_{\rm eff}\right)}{(1+z_s)^5}.
\end{align}
The observed flux, with the lensing magnification accounted for, is then given by an angular integration over the lensed image:
\begin{equation}
    F_\lambda(\lambda, T_{\mathrm{eff}}, g_{\mathrm{eff}}) = \int_I\, S_\lambda\left(\lambda, T_{\rm eff},\,g_{\rm eff}\right)\, \mathrm{d}^2 \Omega.
\end{equation}
The observed flux convolved with a given photometric filter is
\begin{equation}
    \overline{F}_\lambda = \frac{\int_\lambda\, F_\lambda\, f(\lambda)\,\lambda\,\mathrm{d}\lambda}{\int_\lambda\, f(\lambda)\,\lambda\,\mathrm{d}\lambda}.
\end{equation}
where $f(\lambda)$ is the filter throughput. The AB magnitude~\citep{1983ApJ...266..713O} is then computed as
\begin{align}
    m_{\rm AB} = -2.5\,\log_{10}(\overline{F}_\lambda\,\lambda_p^2/c) - 48.60,
\end{align}
where $\lambda_p$ is the pivot wavelength of the filter.

\subsection{Viewing Angles}

\begin{figure}[htbp]
\centering
\epsscale{1.1}
\plotone{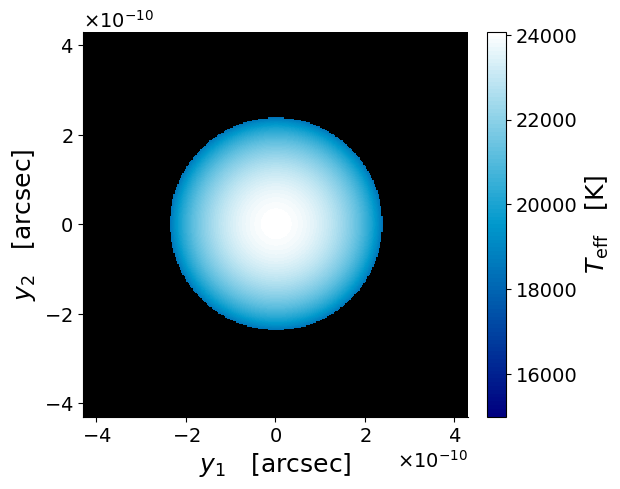}
\plotone{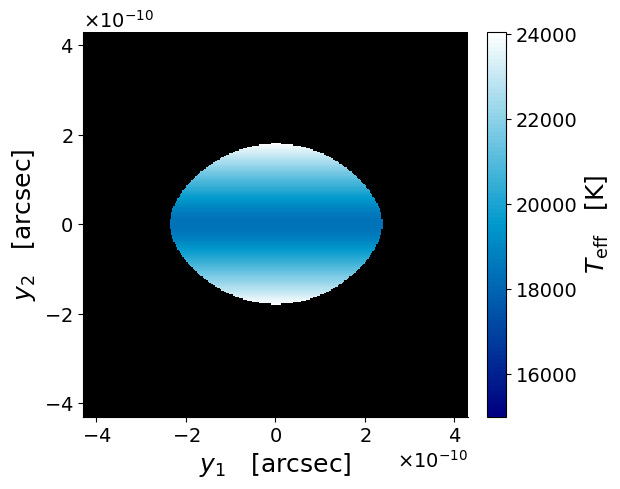}
\plotone{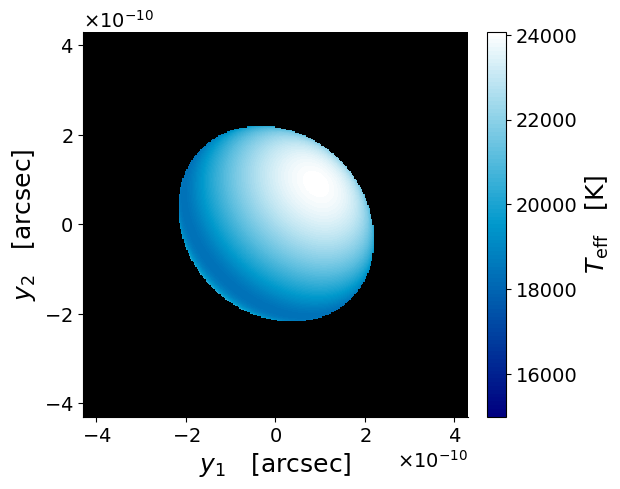}
\caption{Variation of the effective temperature $T_{\rm eff}$ across the surface of a rapidly rotating star with $\omega=0.8$ on the source plane. From top to bottom we show three perspectives: along the rotation axis ($\Theta=0,\,\Phi=0$), in the equatorial plane ($\Theta=\pi/2,\,\Phi=0$), and at an inclined angle ($\Theta=\pi/4,\,\Phi=\pi/4$). The star has an equatorial radius $R_e=83.25\,R_\odot$, a bolometric luminosity $L=10^{6}\,L_\odot$, and a stellar mass $M=60\,M_\odot$, corresponding to an evolved blue supergiant at $Z=0.2\,Z_\odot$. We use an angular diameter distance $D_s = 1.65\,$Gpc corresponding to a source redshift $z_s=1$.
\label{fig:P_T}}
\end{figure}

For a rotating star, the observed surface brightness profile depends on the orientation of the star with respect to the line of sight, which is random. We parameterize this using two of the three Euler angles: $\Theta$ is the angle between the line of sight and the star's rotation axis, and $\Phi$ is the azimuthal position of the rotation axis when projected onto the plane of the sky. On the source plane, the surface brightness profile is simply related by rotations in the plane of the sky for situations with the same $\Theta$ but different $\Phi$ values. \reffig{P_T} shows an example of how the surface brightness of a rotating star varies with the observer's perspective. The angle $\Phi$ is relevant when the star is lensed because the multifilter photometric outcome of the time-varying differential magnification effect during a caustic-crossing event depends on the orientation of the star in the sky plane relative to the direction of the lensing caustic.

\begin{figure}[htbp]
\centering
\epsscale{1.1}
\plotone{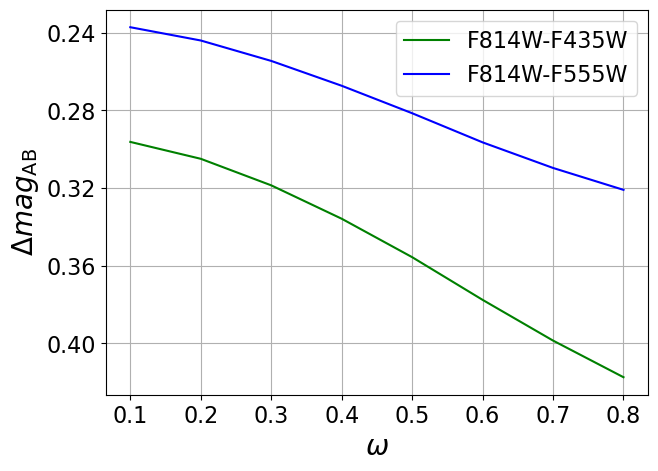}
\plotone{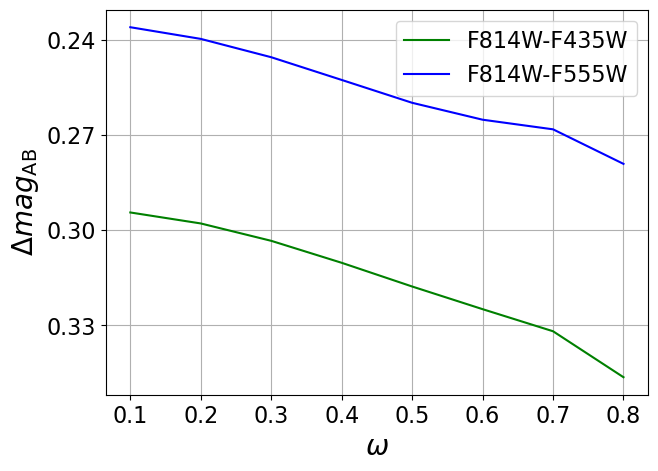}
\plotone{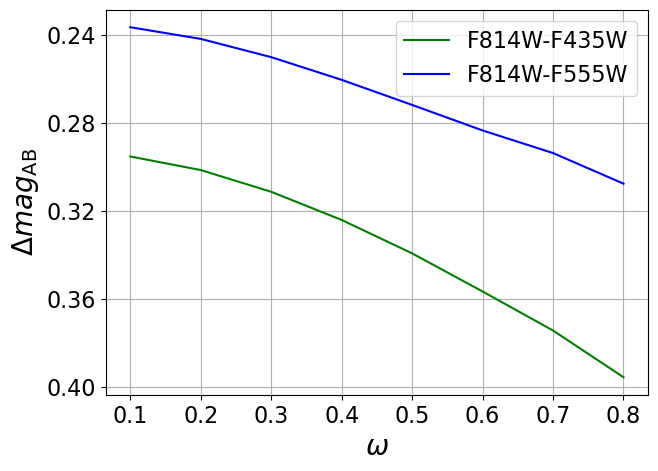}
\caption{Variation in broadband colors as a function of the reduced angular velocity $\omega$ for an unlensed star, which is the same one as in Figure 2 (except for the varying $\omega$). From top to bottom, curves are shown for three perspectives: along the rotation axis ($\Theta=0,\,\Phi=0$), in the equatorial plane ($\Theta=\pi/2,\,\Phi=0$), and at an inclined angle ($\Theta=\pi/4,\,\Phi=\pi/4$).
\label{fig:colors}}
\end{figure}

In Figure~\ref{fig:colors}, we consider the same example star as in Figure~\ref{fig:P_T}. We vary the dimensionless surface rotation $\omega$ while keeping other intrinsic parameters fixed. Broadband colors, corresponding to three promising HST filters (F435W, F555W, and F814W) with which the star may be observed, are shown as a function of $\omega$ and along different viewing angles. In this example, there are only up to $\sim 0.1\,$mag subtle color differences between a nonrotating star and a rapidly rotating star, in the absence of differential magnification. As we will see later, the detailed shape of the flux light curve during the caustic crossing (as a result of time-dependent differential magnification of the photosphere), rather than the color variation, provides the key information to distinguish between different $\omega$ values.

\subsection{Summary of Stellar Parameters}

In summary, we characterize a rotating source star with four intrinsic parameters: equatorial radius $R_e$, bolometric luminosity $L$, mass $M$, and the dimensionless rotation rate $0 \leqslant \omega < 1$; two external parameters describe the viewing angle: the inclination $0 \leqslant \Theta \leqslant \pi$ and the azimuthal angle $0 \leqslant \Phi < 2\pi$.

\section{Microcaustic Crossing} 
\label{sec3}

The situation of the highest observational promise is when a source star approaches a portion of a microcaustic cast by intracluster microlenses. Around this time, the star appears the brightest owing to a pair of highly magnified microimages. This lensing situation can be well described by a universal fold model applied to the microcaustic vicinity~\citep{BlandfordNarayan1986, SchneiderEhlersFalco1992textbook}. In this section, we discuss this fold model in the context of this work.

In geometrical optics and under the thin-lens approximation, the angular coordinates $\Vec{x}$ on the image plane are mapped to the angular coordinates $\Vec{y}$ on the source plane through the ray equation
\begin{equation}
    \Vec{y} = \Vec{x} - \Vec{\alpha}(\Vec{x}),
\end{equation}
where $\Vec{\alpha}(\Vec{x}) = \Vec{\nabla} \psi(\Vec{x})$ is the deflecting angle and $\psi(\Vec{x})$ is the lensing potential.

The lensing Jacobian matrix takes the following form:
\begin{equation}
\begin{split}
    \Vec{A}(\Vec{x}) &\equiv {\partial\, \Vec{y}(\Vec{x})}/{\partial\, \Vec{x}} \\
    & = \left(
    \begin{array}{cc}
        1-\kappa(\Vec{x})-\lambda(\Vec{x}) & - \eta(\Vec{x}) \\
        - \eta(\Vec{x}) & 1-\kappa(\Vec{x})+\lambda(\Vec{x})
    \end{array}
    \right).
\end{split}
\end{equation}
Following the notation in \cite{Venumadhav2017CausticMicrolensing}, we can Taylor-expand the quantities,
\begin{align}
    \kappa(\Vec{x}) &= \kappa_0 + (\Vec{\nabla}\kappa)_0\cdot\Vec{x} + \mathcal{O}(\Vec{x}^2), \\
    \lambda(\Vec{x}) &= 1 - \kappa_0 + (\Vec{\nabla}\lambda)_0\cdot\Vec{x} + \mathcal{O}(\Vec{x}^2), \\
    \eta(\Vec{x}) &= (\Vec{\nabla}\eta)_0\cdot\Vec{x} + \mathcal{O}(\Vec{x}^2),
\end{align}
In addition to the convergence parameter $\kappa_0$, the fold is uniquely characterized by the gradient vector
\begin{equation}
\label{eq:dstar}
    \Vec{d}_\star = -(\Vec{\nabla}\kappa)_0-(\Vec{\nabla}\lambda)_0.
\end{equation}
The lensing Jacobian matrix in the caustic vicinity is
\begin{equation}
\begin{split}
    \Vec{A}(\Vec{x}) &=
    \left(
    \begin{array}{cc}
        (-(\Vec{\nabla}\kappa)_0-(\Vec{\nabla}\lambda)_0)\cdot\Vec{x} & -(\Vec{\nabla}\eta)_0\cdot\Vec{x}\\
        -(\Vec{\nabla}\eta)_0\cdot\Vec{x} & 2\,(1-\kappa_0)
    \end{array}
    \right) \\
    &\simeq 
    \left(
    \begin{array}{cc}
    \Vec{d}_\star\cdot\Vec{x} & 0 \\
    0 & 2\,(1-\kappa_0)
    \end{array}
    \right).    
\end{split}
\end{equation}
The ray deflection $\Vec{\alpha}(\Vec{x}) = (\alpha_1, \alpha_2)$ is given by
\begin{align}
    \alpha_1 &= (1-\frac{1}{2}\,d_{\star\,1}\,x_1 - d_{\star\,2}\,x_2)\,x_1, \\
    \alpha_2 &= (2\,\kappa_0 - 1)\,x_2.
\end{align}

The parameter $\kappa_0$ is set by the macroscopic lens surface mass density over roughly arcsecond scale around the magnified star and can be determined to a decent precision with galaxy cluster mass modeling~\citep{Venumadhav2017CausticMicrolensing,Kelly2018NatAsM1149,Dai2020SGASJ1226}. By contrast, the relevant vector $\Vec{d}_\star$ at each microcaustic crossing is set by the microlenses and has a random value. In the region where the macro critical curve is fully disrupted by microlenses, $\Vec{d}_\star$ has a characteristic magnitude~\citep{Venumadhav2017CausticMicrolensing}
\begin{equation}
    |\Vec{d}_\star| \simeq \kappa_\star^{3/2} / \theta_\star,
\end{equation}
where $\kappa_\star$ is the contribution of intracluster microlenses to the coarse-grained convergence and $\theta_\star$ is the typical angular Einstein radius of microlenses~\citep{Venumadhav2017CausticMicrolensing, Diego2018DMUnderMicroscope, Oguri2018CausticCrossing}. In practice, an informative prior can be supplied for $\Vec{d}_\star$ based on the knowledge of $\kappa_\star$ and $\theta_\star$, i.e. the abundance and typical masses of microlenses.




The above fold model enables us to calculate the lensed flux using the inverse ray-shooting method given a source star and its source-plane location.

\begin{figure*}[htbp]
\centering
\epsscale{1}
\plotone{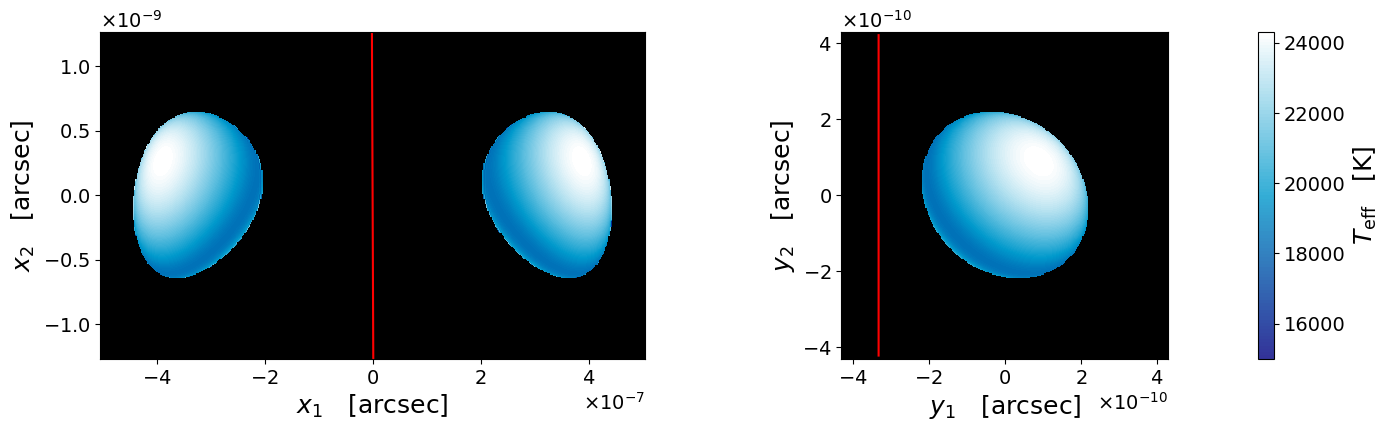}
\plotone{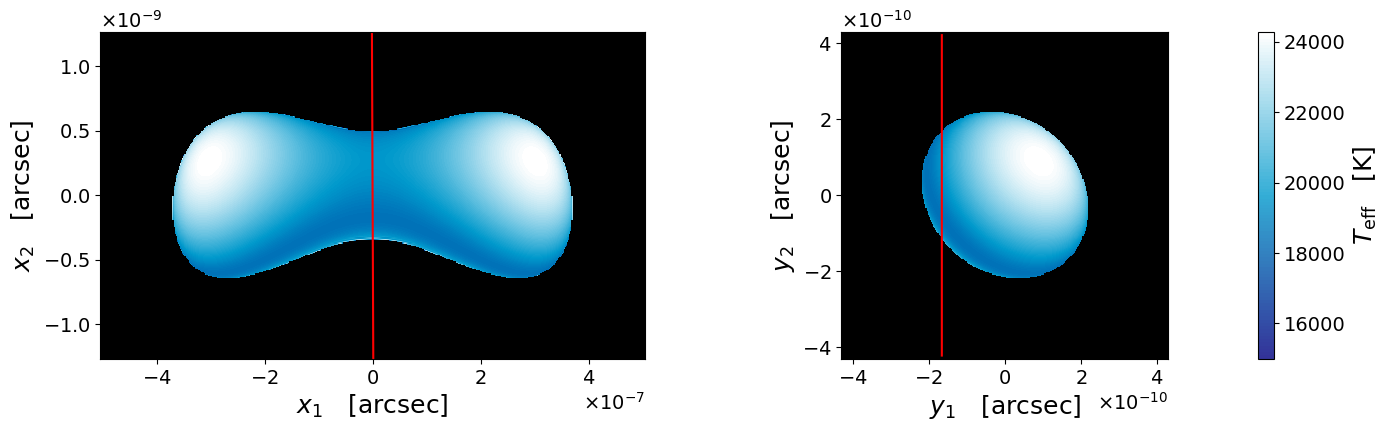}
\plotone{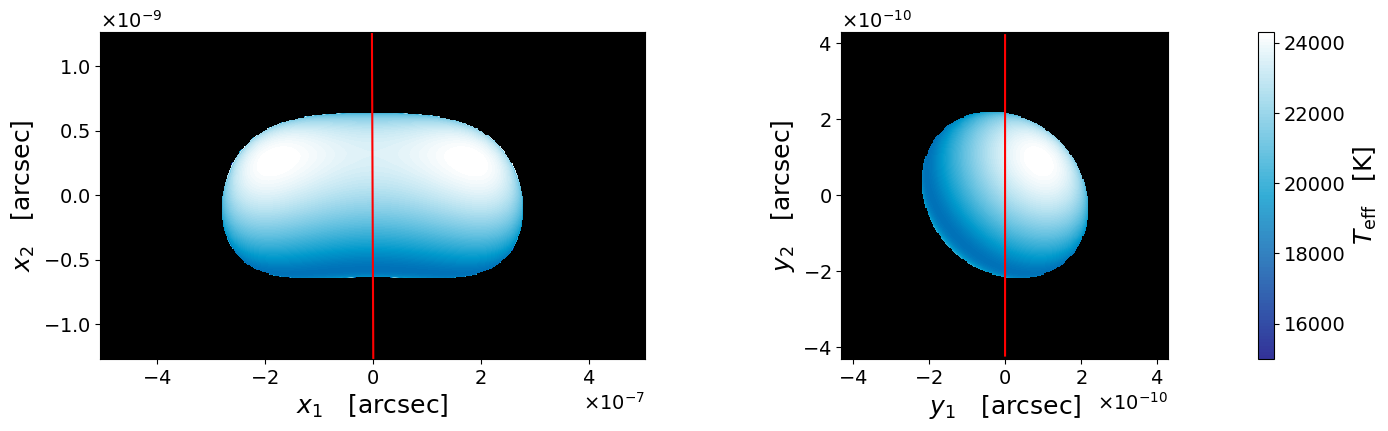}
\plotone{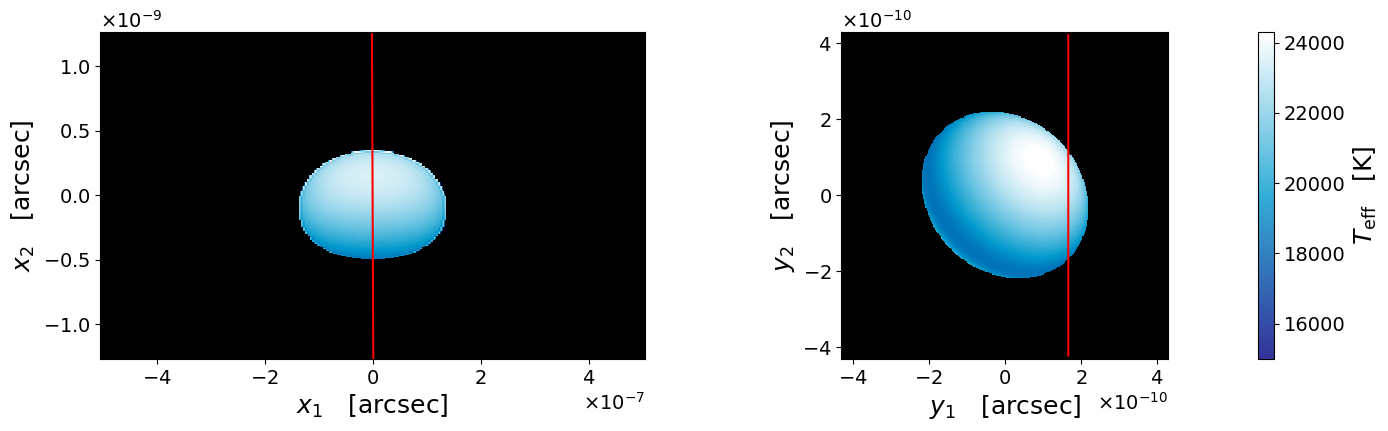}
\caption{A series of snapshots of a caustic crossing event for the same fast rotating star as in \reffig{P_T}, as seen on the image plane (left panels) or on the source plane (right panels), with the red line being the critical curve or the caustic. From top to bottom are four epochs during the process. As the star transits the caustic, the lensed image pair get closer, merge on the critical curve and finally disappear. For the caustic, we set $\kappa_0=0.83$, $\Vec{d}_\star=(5.59, 5.59) \times 10^3\,\mathrm{arcsec}^{-1}$. Note that the panels on the left (showing the image plane) cover a vastly different angular scale along the $x_1$-direction compared to along the $x_2$-direction, reflecting that the lensed images are highly elongated along the $x_1$-direction.
\label{fig:time}}
\end{figure*}

\reffig{time} shows snapshots of a lensed rotating star on the source plane and on the image plane at four different epochs during the microcaustic crossing. As the star approaches the caustic, a pair of images become increasingly bright and stretched. The image pair merge when the source star touches the caustic. The merged image shrinks and eventually disappears as the source star hides on the other side of the caustic.

The motion of the star sets the timescale of the caustic-crossing light curve. To the extent that the curvature radius of the caustic is large compared to the stellar radius, the relevant velocity component is the one perpendicular to the caustic, corrected for time dilation in the expanding Universe~\citep{1991ApJ...379...94M, Venumadhav2017CausticMicrolensing},
\begin{align}
    v_t = \left|\left(\frac{\Vec{v}_s}{1+z_s} - \frac{D_s}{D_l}\,\frac{\Vec{v}_l}{1+z_l}\right) \cdot \hat{\Vec{s}}\right|
\end{align}
Here $\Vec{v}_s$ and $\Vec{v}_l$ are the proper velocities of the source and the lens (relative to the Earth), $\hat{\Vec{s}}$ is a unit vector in the sky perpendicular to the local caustic, $z_s$ and $z_l$ are the source redshift and lens redshift, and $D_s$ and $D_l$ are the angular diameter distances to the source and to the lens, respectively. For a given highly magnified star, $z_l$ and $z_s$ are often both known, and so are $D_s$ and $D_l$. For each microcaustic crossing event, the value of $v_t$ is not known. We therefore include it as a free parameter in light-curve fitting. Finally, the exact time of caustic crossing $t_{\rm cr}$ is unknown and has to be another free parameter.

\section{Parameter Degeneracy}
\label{sec4}


The surface temperature $T_{\rm eff}$ is mainly constrained by the multiband colors. The maximal observed fluxes constrain the luminosity $L$ multiplied by the peak magnification $\mu_{\rm pk}$, i.e. $\mu_{\rm pk}\,L \simeq \mu_{\rm pk}\,4\pi\,R^2_e\,T^4_{\rm eff}$. The value of $\mu_{\rm pk}$ is unknown; it is approximately given by $\mu_{\rm pk} \sim (d_\star\,R_e/D_s)^{-1/2}$. Therefore, the combination $R^{3/2}_e/d^{1/2}_\star$ is well measured. It follows that the logarithmic uncertainty in $R_e$ is $1/3$ of the uncertainty in $d_\star$ and the logarithmic uncertainty in $L$ is $2/3$ of the uncertainty in $d_\star$. The combination $R_e/v_t$ can be inferred from the timescale of peak magnification, and hence the logarithmic uncertainty in $v_t$ is $1/3$ that in $d_\star$. These conclusions are confirmed by our mock parameter inferences (see \reffig{corner04} and \reffig{corner01}).

Combining \refeq{F} and \refeq{geff} and taking into account the dimensionless equations \refeq{shape} and \refeq{shapeEq}, we see that if two stars have identical $L$, $R_e$, $\omega$ and $Z$ but different $M$ values, they have the same surface shape but only exhibit subtle differences in their SEDs that reflect the effect of $g_{\rm eff}$. This provides the unique information to infer $M$. Since this information cannot be efficiently extracted through wide-filter photometry, we expect the inference precision for $M$ to be poor (see \reffig{corner01} and \reffig{corner04}) and perhaps prone to uncertainty in the modeling of spectral line features in the SED. 

In this study, we treat the macro convergence $\kappa_0$ as a known parameter. In reality, its inference from galaxy cluster lens modeling is subject to uncertainty. Since $\kappa_0$ directly impacts only the magnification perpendicular to the direction of image elongation, we expect that if uncertainty in $\kappa_0$ is included it will add to the degeneracy involving $L$, $R_e$, $\Vec{d}_\star$ and $v_t$, for which uncertainty is in any case dominated by the significant uncertainty in $\Vec{d}_\star$. We therefore expect that the uncertainty in $\kappa_0$ does not impact the inference of $\omega$.

We note that for the most massive stars, especially the evolved supergiants, the bolometric luminosity $L$ approaches the Eddington limit $L_{\rm Edd}=4\pi\,G\,M\,c/\kappa_T$, where $\kappa_T$ is the opacity of Thomson scattering. The Eddington ratio $L/L_{\rm Edd}$ is not precisely unity, and the precise value depends on stellar structure and wind models. Combining constraints from direct observation of caustic crossing and massive star models may enable a significantly improved mass measurement.

While the inference precisions for multiple intrinsic parameters are fundamentally limited by $d_\star$, which is not directly measurable, it is possible to derive a reliable and informative prior on $d_\star$. For a given highly magnified star on a given caustic-crossing giant arc, the parameters of the local macro caustic can be obtained through lens modeling, while both the abundance and mass distribution of the microlenses can be obtained from photometric analysis of the intracluster light (e.g., \cite{Kelly2018NatAsM1149, Kaurov2019LensedStarM0416, Dai2020SGASJ1226}). With such knowledge, an accurate prior for $d_\star$ can be derived, for instance, by numerically simulating random microlensing realizations. We crudely estimate that the fractional RMS in $d_\star$ will be $\simeq 0.35\,$dex, a number we use to set our priors for $d_\star$ in the next section. For this value, we expect $\sim 0.1\,$dex uncertainty in $R_e$ and $v_t$ and $\sim 0.2\,$dex uncertainty in $L$, provided that errors of photometry do not dominate the error budget.

\section{Mock Parameter Inference}
\label{sec5}

In this section, we present mock parameter inference examples.
We consider two scenarios: one star with a dynamically significant rotation speed $\omega=0.4$, and another star with a slow rotation $\omega=0.1$. In both cases, we set the same values for the other intrinsic and extrinsic parameters (true values and assumed prior distributions summarized in \reftab{table1}), corresponding to an $M=60\,M_\odot$ blue supergiant with a surface temperature $T_{\rm eff}\approx 20,000\,$K and a bolometric luminosity $L=10^6\,L_\odot$ at a metallicity $Z=0.2\,Z_\odot$ and from $z_s=1$.

The mock light curves are the sum of injected light curves and random photometric noise. To collect crucial color information, we consider monitoring in three HST wide filters, F435W, F555W, and F814W, with the added random photometric noise corresponding to the 1$\sigma$ AB magnitude limit $29.0$, $29.0$, and $29.3$ per epoch, respectively. The photometric sensitivities assumed here are realistic for the HST, as they are comparable to the magnitude limits of the Hubble Frontier Fields (HFF) program~\citep{Lotz2017HubbleFrontierFields} when the exposure time per observation is rescaled to be about 1 hr per filter (compatible with the cadence $\lesssim 4\,$hr for three filters). JWST will be more sensitive but can only observe $\lambda>0.7\,\mu\mathrm{m}$. The logarithmic likelihood function is taken to be minus one-half of the overall fitting $\chi^2$, under the assumption that individual photometric measurements at different epochs or in different filters have uncorrelated measurement errors.

\begin{figure*}[htbp]
\centering
\epsscale{1}
\plotone{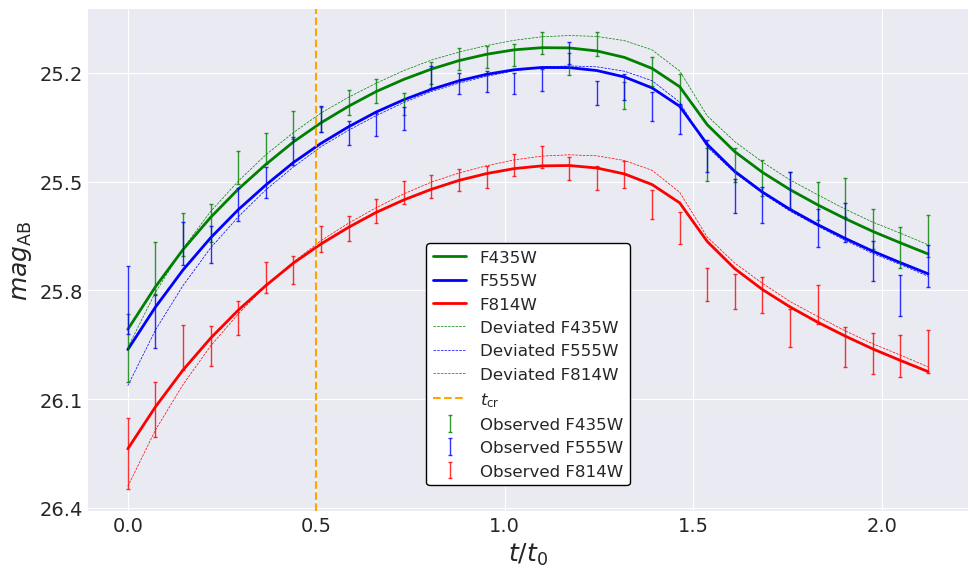}
\plotone{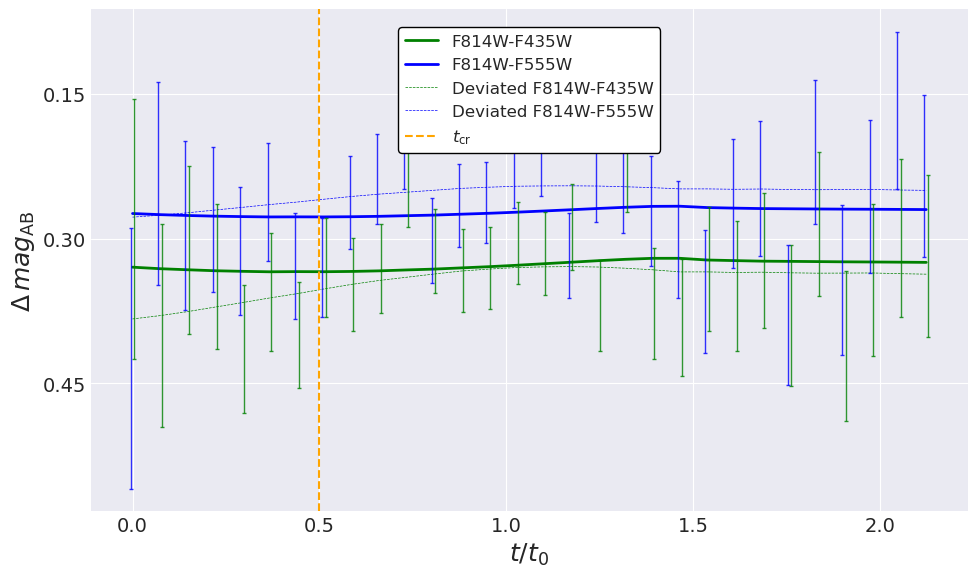}
\caption{Mock multiband light curves (top panel) and the associated subtle color variability (bottom panel) for a lensed rotating star with $\omega=0.4$ crossing a microcaustic. The center of the star passes the caustic at $t=t_{\mathrm{cr}}$ (vertical yellow dashed lines), and the temporal unit $t_0$ is the time it takes for the star to move a distance of its diameter $2R_e$ at $v_t=8.2 \times 10^{2}\,\mathrm{km\,s^{-1}}$, which is $t_0 = 39.25\,$hr in this case. The observation cadence in each filter is about $0.1\,t_0 \approx 4\,$hr. Injected model parameters can be found in Table \ref{tab:table1}. The deviated light curves and the associated color variability are also shown (dashed lines), in which the parameter $\omega=0.7$. In this example, the surface rotation rate $\omega$ is constrained mainly by the detailed shapes of the flux light curves (top panel) than by the subtle color variations (bottom panel).
\label{fig:lightcurves}}
\end{figure*}

In Figure \ref{fig:lightcurves} we show the example mock light curves, color variability, and the best model fit. The peak magnitudes in the optical are $25.2$--$25.5$. These are comparable to the brightest magnitudes observed in the light curve of the first discovered highly magnified star in \cite{Kelly2018NatAsM1149}. Given that that star is from $z_s=1.5$, the observability represented by our mock event is not exaggerated.

In Figure~\ref{fig:lightcurves}, we also show model light curves (dashed lines) corresponding to a different parameter set that yields a fitting $\chi^2$ that is approximately 55 units worse than the best-fit value.
From both panels of Figure~\ref{fig:lightcurves}, we conclude that it is the detailed shapes of the light curves during caustic crossing, more than the color changes, that carry the dominant information for constraining the surface rotation rate.

In the Appendix \ref{app:cornerplots}, we show posterior distributions that result from the mock parameter inferences we perform. Figure \ref{fig:corner04} shows the case of the fast-rotating star with $\omega=0.4$. The true value of $\omega$ is recovered with a fractional error of about $35\%$. The fractional uncertainties of $R_e$, $L$ and $v_t$ scale with the uncertainty in $d_{\star 1}$, the component of $\Vec{d}_\star$ along the degenerate direction, which is the limiting factor and is set by the prior we adopt for $\Vec{d}_\star$. This is reflected in the degeneracy between $d_{\star 1}$ and any of $R_e$, $L$ and $v_t$ as explained in Section \ref{sec4}. No significant degeneracy is seen between $d_{\star 2}$, the other component of $\Vec{d}_\star$, and intrinsic stellar parameters. We are able to distinguish between the nearly pole-on or moderately inclined perspective and a perspective in the equatorial plane, and we derive a decent constraint on the position angle $\Phi$.

For a comparison, Figure \ref{fig:corner01} shows the posterior distributions derived from observing the identical source star but with a dynamically unimportant rotation $\omega=0.1$. A similar absolute uncertainty is inferred for $\omega$. The constraints on the viewing perspective are significantly weaker, as expected for a star with much weaker variation of temperature across its surface. The fractional uncertainties in $R_e$, $L$ and $v_t$ are similar to the $\omega=0.4$ case as is set by the prior used for $\Vec{d}_\star$. 

It is worth mentioning the possible issue of nonaxisymmetric features on the surfaces of rotating massive stars. For instance, early-type B stars have been shown to display rotational variations with amplitudes of up to $0.01\,$mag~\citep{balona2022rotation}. Given our assumed photometric precision 0.03 mag under realistic observing conditions, this effect is not likely to have a significant impact on the parameter inference results.



\section{Discussion}
\label{sec6}

\subsection{Surface Rotation}

An important finding from our study is the ability to infer dimensionless surface rotation. The most massive stars, relevant for extragalactic caustic crossings, are expected to have significant rotation when they start on the main sequence~\citep{MaederMeynet2012RevModPhysReview}. For an $M=50\,M_\odot$ zero-age main-sequence (ZAMS) star, the breakup velocity can be $v_{\rm eq}=800$--$900\,{\rm km}\,{\rm s^{-1}}$ at $Z=0.1$--$1\,Z_\odot$ and higher at lower $Z$~\citep{Ekstrom2008OriginBeStars}. Based on observed examples, $v_i=200$--$500\,{\rm km}\,{\rm s^{-1}}$ are considered plausible values for the ZAMS surface rotation velocity. Dynamically significant initial values $\omega=\mathcal{O}(1)$ are hence expected for massive stars. 

However, the most massive stars on the main sequence (O stars) are currently not the most promising target highly magnified stars observationally (from $z_s=0.7$--$1.5$). Because of their high $T_{\rm eff} \gtrsim 25,000\,$K, they do not appear the brightest in HST or JWST bands~\citep{Kaurov2019LensedStarM0416}. Indeed, the first detected highly magnified star is a cool supergiant with $T_{\rm eff}=11,000$--$14,000\,$K~\citep{Kelly2018NatAsM1149}. The magnified star reported by \cite{Chen2019LensedStarM0416} appears to be photometrically fit by a cool supergiant with $T_{\rm eff}=13600\,$K, while a dust-reddened, hot O supergiant with $T_{\rm eff} \simeq 40,000\,$K also fits photometry decently but is deemed much less probable.

Except under metal-poor conditions, angular momentum loss from stellar wind is expected to significantly reduce surface rotation through the main-sequence lifetime. It raises the question whether warm or cool supergiants (B supergiants) as evolved massive stars may have dynamically significant surface rotation. In the Galaxy, some examples of rotating blue supergiants are known. Sher 25 in the H II region NGC 3603 has $M \simeq 40\,M_\odot$, $T_{\rm eff}=22,000\,$K and $\omega\simeq 0.2$~\citep{Hendry2008Sher25nebula}. Blue supergiant SBW1 has $T_{\rm eff}=21,000\,$K and $\omega\simeq 0.2$~\citep{Smith2017SBW1}. Another blue supergiant HD 168625 has $R=105\,R_\odot$ and $T_{\rm eff}=14000\,$K (perhaps most similar to the highly magnified star reported in \cite{Kelly2018NatAsM1149}) and has a surface rotation $\omega\simeq 0.27$~\citep{Mahy2016HD168607}. The significant surface rotations of these example blue supergiants are also evidenced from their associated ring-shaped nebulae. For this reason, SBW1 is considered an analog of the SN 1987A progenitor~\citep{Smith2017SBW1}. Recently, it has been found in the IACOB spectroscopic survey that Galactic blue supergiants mostly have $v\,\sin\Theta < 70\,{\rm km}/{\rm s}$ for $T_{\rm eff} < 21,000\,$K, except for only a handful of stars with $v\,\sin\Theta = 100$--$150\,{\rm km}\,{\rm s^{-1}}$~\citep{deBurgos2023arXiv231200241D}.

It might be that there are rotating blue supergiants with higher $\omega$ in lower-metallicity galaxies given that wind-driven spin-down from initial rotation is less efficient~\citep{MeynetMaeder2002StellarEvolutionWithRotationIII}. Additionally, critical surface rotation can be reached for a blue supergiant that has recently contracted from a red supergiant phase~\citep{HegerLanger1998, MeynetMaeder2000StellarEvolutionWithRotation}. It is also plausible that tides in tight massive star binaries can make fast-rotating supergiant stars. In any case, the example rotating stars we consider here with $\omega=0.1$ and $0.4$ bracket the likely range of $\omega$ for rotating blue supergiants with $T_{\rm eff} < 21,000\,$K.

\subsection{Extensions of Model}
\label{sec:extmodel}

Our model can be extended to account for other real-world effects. In practice, dust reddening has to be accounted for to correctly interpret the observed colors~\citep{Kelly2018NatAsM1149}. The dust reddening effect is not included in the model we have presented here, but it would be straightforward to introduce one more free parameter to the model provided that the reddening curve is known. Such a new parameter would not be degenerate with $\omega$ because flux variations due to differential magnification are time-dependent signals, while dust reddening is time independent.

There can be other microimages than the two that are highly magnified around the time of caustic crossing~\citep{Venumadhav2017CausticMicrolensing}. Those additional images are significantly fainter and have nearly constant fluxes over the short diameter crossing time $t_0=2\,R_e/v_t$. Therefore, one additive flux constant per filter can be introduced as additional free parameters, and we do not expect them to be degenerate with the intrinsic parameters.

We have not accounted for limb darkening. This effect has been studied for nonrotating OB stars~\citep{WadeRucinski1985, Claret2004NonlinearLTE, Howarth2011a, Reeve2016LimbDarkeningNonLTEHotStars} and is expected for rotating stars too. While it is not the intention of this work to incorporate the limb-darkening effect in addition to gravity darkening, our modeling framework can be generalized to account for it. In that case, the emergent SED at any given point on the stellar surface not only is a function of $Z$ and the local $(T_{\rm eff}, g_{\rm eff})$ but also depends on $\mu$, the cosine of the oblique angle (the \textit{}angle between the line of sight and the local surface normal). The dependence on $\mu$ can be derived from radiation transfer calculations for stellar atmosphere models. \cite{Reeve2016LimbDarkeningNonLTEHotStars} pointed out that for OB stars limb darkening is moderately sensitive to surface gravity, which may provide new information for improving mass inference. In terms of observable effects, limb darkening is partially degenerate with gravity darkening for $|\cos\Theta|\approx 1$ but not so for $|\cos\Theta|\approx 0$. However, if limb darkening can be predicted by radiation transfer calculations as a function of $(\omega,\,L,\,M,\,Z)$, then it merely complicates the SED profile model but is not a free parameter. 

Another obvious direction to extend this work would be to construct a model for binary systems. Massive stars are commonly found to be in binary systems, and the fraction of interacting binaries can be higher than $70\%$~\citep{Sana2012Sci...337..444S}. If the binary separation is large enough (say, larger than several au), the surface brightness profiles of the two stars may each be well described by single-star models with appropriate parameters. In this case, the orbital period is expected to be much longer than the diameter crossing time, and the light curves would be trivial superpositions of two caustic crossings, with a time lag and a doubling of parameters. The more complicated situations could involve tidally induced surface deformations or ongoing mass transfer. The caustic-crossing phenomenon in principle would enable us to ``see'' the details.

If multiple microcaustic crossings of the same source star are observed, it would be possible to tighten up parameter inference by jointly analyzing light curves of all crossings. All crossings share the same model parameters except for $v_t$ and $\Vec{d}_\star$, and a reasonable prior distribution for $\Vec{d}_\star$ specific to each highly magnified star can be theoretically derived. An in-depth study of this interesting method may be pursued in future work.

\begin{table*}[t]
\centering
\begin{tabular}{lllll}
\hline
Model Parameter & Symbol & True Value(s) & Prior Distribution Type & Range or Prior Parameters \\
\hline
Dimensionless rotation & $\omega$ & 0.1, 0.4 & Uniform & $[0, 0.8]$ \\
Equatorial radius & $R_e\,(R_\odot)$ & $83.25$ & Log-uniform & $[10, 1000]$ \\
Bolometric luminosity & $L\,(L_\odot)$ & $10^{6}$ & Log-uniform & $[10^5, 10^7]$ \\
Stellar mass & $M\,(M_\odot)$ & $60$ & Log-uniform & $[10, 1000]$ \\
Inclination & $\Theta$ & $\pi/4$ & Cosine-uniform & $[0, \pi/2]$ \\
Position Angle & $\Phi$ & $\pi/4$ & Uniform & $[0, \pi/2]$ \\
Microcaustic strength & $\Vec{d_\star}$ (arcsec$^{-1}$) & (5.59, 5.59)$\times 10^{3}$ & Log-normal & $\mu=\mathrm{lg}\,(5.59 \times 10^{3}), \,\sigma=0.35$ \\
Effective transverse velocity & $v_t\,({\rm km}\,{\rm s}^{-1})$ & $8.2 \times 10^{2}$ & Uniform & $[300, 3000]$ \\
Epoch of caustic crossing & $t_\mathrm{cr}\,({\rm hr})$ & $19.63$ & Uniform & $[0, 39.25]$ \\
\hline
\end{tabular}
\caption{Model parameters and the prior distributions we use in mock parameter inference. The physically allowed range for the dimensionless rotation parameter $\omega$ is $0 \leqslant \omega < 1$. We consider a prior range between 0 and 0.8 because numerical difficulty arises for $\omega$ values near breakup ($\omega \approx 1$) and the mock stars we choose are not rotating that fast in any case. Due to reflection symmetries it is sufficient to sample one quadrant of the position angle range, $0 \leqslant \Phi \leqslant \pi/2$. For $\Vec{d}_\star$, the same lognormal prior applies to both Cartesian components $d_{\star 1}$ and $d_{\star 2}$. The true values of the model parameters we choose are not fine-tuned.
}
\label{tab:table1}
\end{table*}

\section{Conclusion}
\label{sec7}

We have studied a method to measure stellar parameters of individual highly magnified stars in caustic-overlapping lensed galaxies at cosmological redshifts. The method exploits light variability as a result of the differential magnification effect when the source stars transit microlensing caustics cast by intracluster stars.

We have constructed an analytic SED model that quantifies the axisymmetric surface brightness profile induced by the gravity-darkening effect for rotating O/B stars (Section \ref{sec2}). The SED model specifically builds on the TLUSTY SED templates for nonrotating O/B stars, although our framework is applicable to general stellar SED templates.
Gravitational lensing in the vicinity of the microcaustic is modeled, at least near the epoch of caustic crossing, by a simple fold (Section \ref{sec3}). Combining the model SED profile and the lensing effect, we have introduced a light-curve model parameterized by intrinsic stellar parameters and extrinsic geometric and kinematic parameters.
Based on the parameterized light-curve model, we have developed a general code for performing parameter inference with multifilter light curves.

We have demonstrated the method by performing mock parameter inference under observational conditions that are realistic for the HST (three wide filters, 1$\sigma$ magnitude limit per epoch $\sim 29$, and at a cadence of a few hours over a few days), for a blue supergiant source star at $z_s=1$ (Section \ref{sec5}). Our results suggest that the surface rotation velocity relative to the breakup value can be measured to an error $\sigma_\omega = 0.1$--$0.2$, without a geometric degeneracy, while other intrinsic parameters such as equatorial radius $R_e$ and bolometric luminosity $L$ are limited by uncertainty about the microcaustic strength and are likely to be determined to $0.1$ and $0.2\,$dex (1$\sigma$), respectively. Mass inference is less constraining but could be improved with a careful treatment of limb darkening. 

While there is much room for further improvement in the modeling of the surface brightness profile, this first study suggests a new opportunity to survey the individual properties of the most massive and luminous stars in distant high-$z$ galaxies, which may advance our understanding of massive star formation and evolution. While we have not examined particularly lensed stars in higher-$z$ lensed galaxies (say, $z>2$), that case could be even more interesting, since there would be increased chance of seeing stars in metal-poor systems~\citep{Welch2022EarendelNature} and cosmological redshifting should make hotter (and more common) massive stars observationally more accessible. Observational prospects for this might arise in the future.

In the situation of recurring microlensing brightening events in the same lensed galaxy, the feasibility to significantly constrain intrinsic and extrinsic parameters can help us determine which caustic-crossing events are associated with the same source star. This knowledge would be useful in realizing highly magnified stars as a novel dark matter probe, such as uncovering star-free subgalactic cold dark matter halos through astrometry~\citep{Dai2018Subhalo, Abe2023arXiv231118211A, Williams2023arXiv230406064W} or dense minihalos composed of QCD axion particles as the dark matter~\citep{DaiMiralda2020AxionMinihalo, Xiao2021AxionMinihalo}.

\section*{acknowledgments}

The authors would like to express thanks to the Berkeley Physics International Education (BPIE) program through which X.H. visited UC Berkeley as an undergraduate student and during which this research project was initiated. The authors thank Maude Gull for useful discussions. L.D. acknowledges research grant support from the Alfred P. Sloan Foundation (award No. FG-2021-16495) and the support of the Frank and Karen Dabby STEM Fund in the Society of Hellman Fellows.

\appendix

\section{Posterior distributions of mock parameter inferences}
\label{app:cornerplots}

\begin{figure*}[htbp]
\centering
\epsscale{1.2}
\plotone{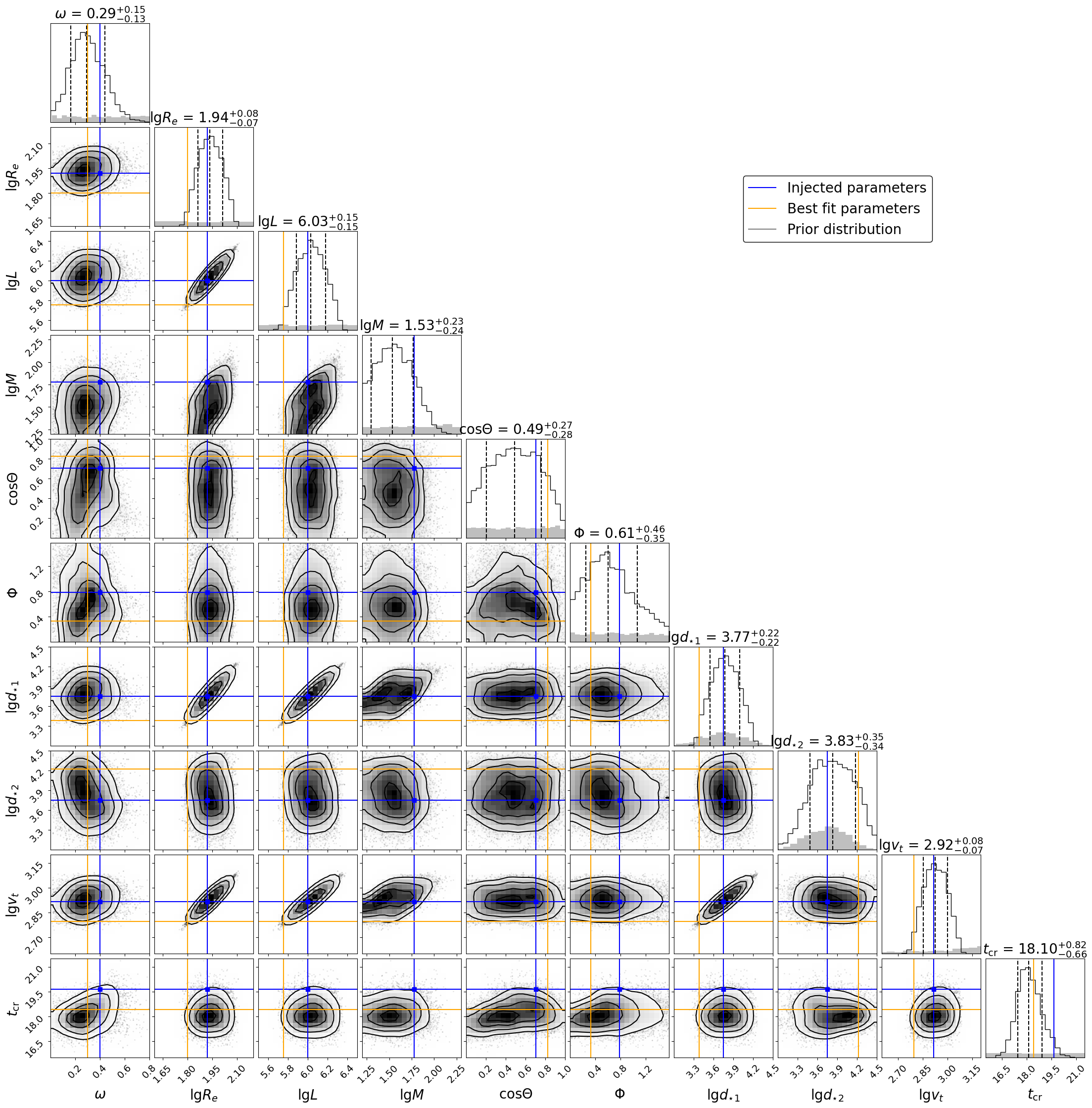}
\caption{1D and 2D marginalized posterior distributions for the case of rapid surface rotation $\omega=0.4$. The blue and yellow lines mark the true parameters and the maximal likelihood parameters, respectively. 
In the panels showing 1D distributions, the prior distribution is shown by the shaded gray histogram. Among the three vertical dashed lines, the middle one marks the median of the posterior distribution, and the left and right ones mark the 16th and 84th percentiles of the distribution, respectively.
In the panels showing 2D distributions, the three contours, from the innermost to the outermost, enclose the 68.27th, 95.45th, and 99.73th percentiles of the distribution, respectively. We refer readers to Table~\ref{tab:table1} for the units of the dimensionful parameters.
\label{fig:corner04}}
\end{figure*}

\begin{figure*}[htbp]
\centering
\epsscale{1.2}
\plotone{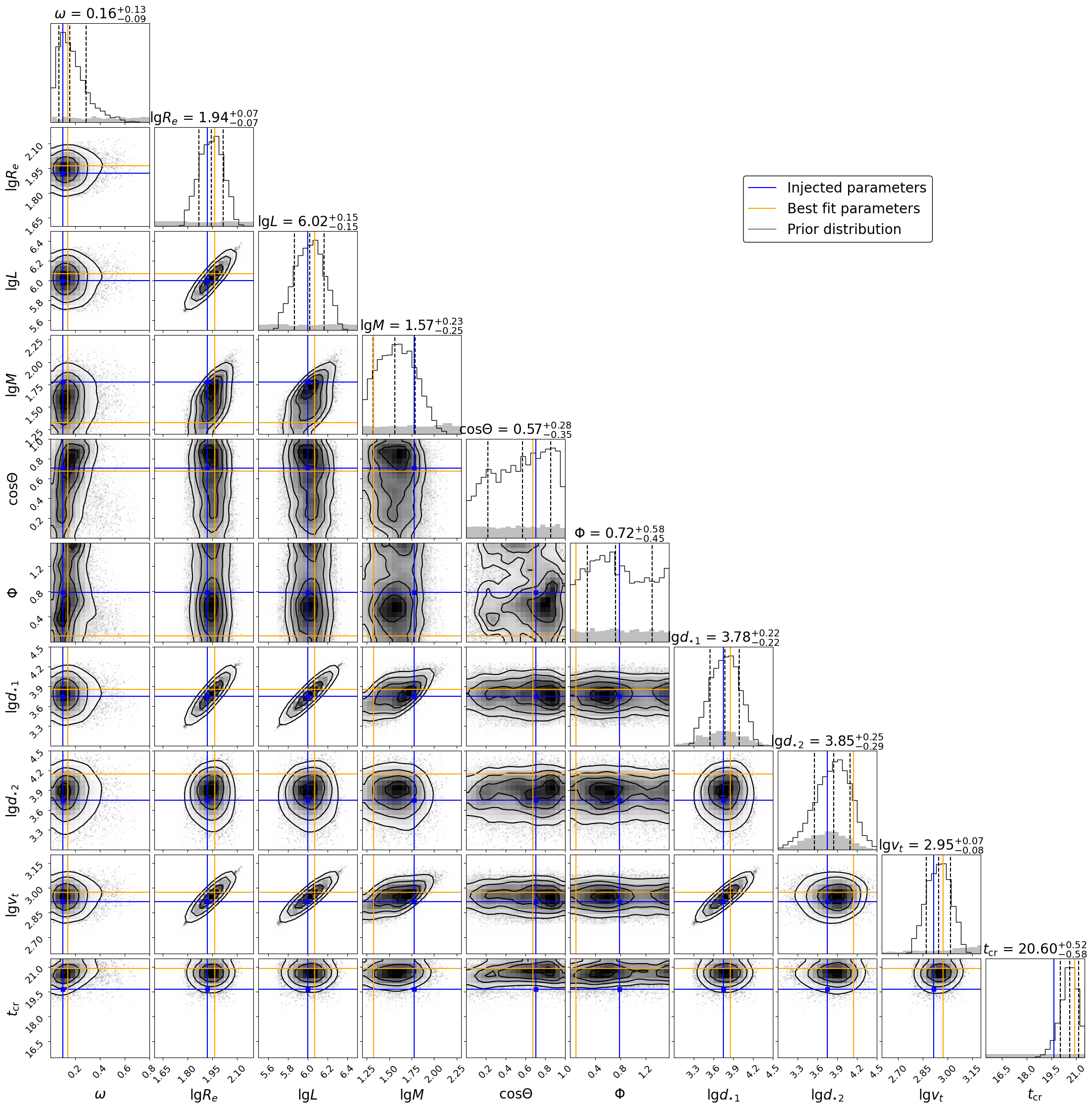}
\caption{Same as Figure \ref{fig:corner04}, but for the case of slower surface rotation $\omega=0.1$.
\label{fig:corner01}}
\end{figure*}

In this appendix, we show the posterior distributions obtained from mock parameter inference runs for the full list of model parameters. Figure \ref{fig:corner04} shows the case of the fast-rotating star with $\omega=0.4$. Figure \ref{fig:corner01} shows the posterior distributions derived from observing the identical source star but with a dynamically unimportant rotation $\omega=0.1$.

\bibliography{references}{}
\bibliographystyle{aasjournal}

\end{document}